\documentclass[twocolumn,showpacs,preprintnumbers,amsmath,amssymb]{revtex4}

\usepackage{graphicx}
\usepackage{dcolumn}
\usepackage{bm}

\def\be{\begin{equation}}
\def\ee{\end{equation}}
\def\f{\frac}
\newcommand{\arr}{\rightarrow}
\def\la{\langle}
\def\ra{\rangle}
\newcommand{\lalg}[1]{\mathfrak{#1}}  
\newcommand{\so}{\lalg{so}}
\newcommand{\su}{\lalg{su}}
\newcommand{\N}{\mathbb{N}}

\begin{document}

\title{Some Remarks on the Semi-Classical Limit of Quantum Gravity}

\author{Etera R. Livine}
\email{elivine@perimeterinstitute.ca}

\affiliation{Perimeter Institute, 31 Caroline Street North, Waterloo, ON, Canada N2L 2Y5}

\date{January 12nd 2005}

\begin{abstract}

One of the most important issues in quantum gravity is to identify its semi-classical regime. First the issue is to define for we mean by a semi-classical theory of quantum gravity, then we would like to use it to extract physical predictions. Writing an effective theory on a flat background is a way to address this problem and I explain how the non-commutative spacetime of deformed special relativity is the natural arena for such considerations. On the other hand, I discuss how the definition of the semi-classical regime can be formulated in a background independent fashion in terms of quantum information and renormalisation of geometry.

\end{abstract}

\maketitle


Building a theory of Quantum Gravity is one of the biggest problems in theoretical physics. In this context, the concept of semi-classicality is particularly relevant in order to make it a physical theory. Indeed, on a practical level, building an effective theory would allow to extract physics and experimental predictions for quantum gravity. In fact, it would even allow us to say something on quantum gravity without having to build the full consistent theory. This is particularly interesting since experiments like GLAST, AUGER and others, should be able to
measure effects due to a quantum gravity regime.

Now, in order to define a semi-classical domain, one should first understand the notion of observers in a quantum gravity context. Then one could construct an effective theory for an observer who can define a regime in which geometry doesn't need to be described as quantum anymore. First, I will describe the different limits of quantum gravity and what I call an effective theory. Then I will show that the non-commutative space-time of deformed special relativity provides an effective theory for quantum gravity. And I will define the semi-classical regime of the theory. I will discuss the necessity of introducing coherent states in a non-commutative spacetime in order to describe a semi-classical notion of spacetime points. Section IV will discuss the interplay of quantum information and quantum gravity. Finally, I will discuss the tools necessary in order to understand the semi-classical regime, or more generally a macroscopic regime, in full background independent quantum gravity such as loop quantum gravity.

\section{Effective Theory and Semi-Classicality}

Quantum Gravity is usually defined as the regime where the Planck quantities become relevant:
\be
l_P=\sqrt{\f{2G\hbar}{c^3}}, \qquad m_P=\sqrt{\f{\hbar c}{2G}},
\ee
where $G$ is the Newton coupling constant for gravity, $c$ the speed of light or highest speed of any signal in flat spacetime, $\hbar$ the Planck constant of the quantum theory.
It is usually believed that gravitational effects (black hole formation) implies that the Planck length $l_P$ is a minimal distance and that the geometry is quantized in quanta of size $l_P$. The relevance of the Planck mass $m_P$, or Planck energy $E_P=m_P c^2$, is more subtle. It can not be a maximal mass since macroscopic systems have a rest energy much larger than $E_P$. Obviously, it can not be a minimal mass either. An easy way to characterize it is as the maximal energy or mass content of a cell of size $l_P$.

More precisely, let us consider a system of characteristic size $L$. Then black holes in general relativity imply a maximal bound on the mass of the system:
\be
M\le M_{max}(L)\equiv \frac{c^2L}{2G}.
\ee
On the other hand, quantum effect implies mass fluctuations of the system of order:
\be
\delta M(L)\equiv \frac{\hbar}{cL}.
\ee
Assuming that general relativity and quantum mechanics still hold at such scale, the Planck scale $l_P$ can be defined as the scale where quantum fluctuations automatically saturate the gravitational bound: $M_{max}(l_P)=\delta M(l_P)$. Then the mass bound is $m_P$. 

Furthermore, there is another quantity relevant to quantum gravity. It is the cosmological constant $\Lambda$, who provides in more local terms an energy density or a local mean curvature. It provides us with a second length scale which we can use to obtain a dimensionless parameter: $z\equiv l_P\sqrt{\Lambda}$. $z$ is usually the quantum deformation parameter in the quantum group approach to quantum gravity. 
In the full theory, we expect $l_P$ to impose a minimal bound on distances while the "cosmological" length $l_C \equiv 1/\sqrt{\Lambda}$ defines a maximal scale. 
Note that, even starting with a vanishing cosmological constant, we would necessarily get $\Lambda\ne 0$ under a coarse-graining flow or more generally a renormalisation group flow.

The different limits of quantum gravity can be defined as $\hbar\arr 0$ (the classical limit), $c\arr\infty$ (the Newtonian limit), $G\arr 0$ (the no-gravity limit or quantum field theory limit), $\Lambda\arr 0$ (which I call the flat spacetime limit). This is usually translated in Planck units into $l_P\arr 0$, $m_P\arr\infty$, $z\arr 0$. However generically one can have more subtle mathematical limits  having some Planck units running to 0 while other ones remain finite. 

\medskip

I define an effective theory in a generic regime when the considered scales are much larger compared to the quantum gravity scales. This still allows quantum effects and general relativistic effects. More operationally, I consider an observer, provided with a resolution $\delta l_{obs}$. Gravity imposes a minimal bound on the observer's resolution, $\delta l\gtrsim l_P$, but the observer doesn't necessarily saturate it. Then there is the scale $L$ of the observed phenomena. As explained above, the two relevant mass (or energy) scales relevant here are a lower bound imposed by quantum effects and related to the resolution $\delta l$ and a maximal bound imposed by general relativity and related to the system scale $L$:
\be
\delta M =\f{\hbar}{c}\f{1}{\delta l}
\le M\le
\f{c^2}{2G}L=M_{max}.
\ee
The natural requirement that $\delta M\le M_{max}$ is equivalent to the assumption $L\,\delta l \ge l_P^2$. In the limit case that $L=\delta l=l_P$, we recover the Planck regime $\delta M=M_{max}=m_P$.

My first assumption for an effective theory of quantum gravity only concerns the observer and will be that $\delta l \ll l_C$. Then, using the equivalence principle of general relativity, I translate the fact that all manifolds are locally flat in an second assumption that the observer will describe the spacetime as being flat. The goal of the effective theory is to describe quantum gravity effects as on that flat spacetime arena.


A semi-classical regime is naturally defined such that quantum fluctuation are not relevant: $\delta M(\delta l) \ll M_{max}(L)$, so that we would be far from a strong quantum gravity regime and black hole formation would be improbable. From this point of view, working out an effective theory doesn't need to work in a semi-classical limit, but that limit will still need to be considered in the derived effective framework.

I propose this criteria to replace different other choices of definition for semi-classicality such as $\hbar\arr 0$, $G\arr 0$, coarse-grained limit $L\gg \delta l$, flat limit $g_{\mu\nu}\sim \eta_{\mu\nu}$, energy scales very large compared to $m_P$, interaction energy scales much smaller than $m_P$.

\section{Deformed Special Relativity}

I would like to show that deformed special relativity\cite{dsrintro1,dsrintro2} provides an arena in which to build an effective theory for quantum gravity. These results are explained in details in \cite{dsreffect}. I consider the cosmological constant $\Lambda$ as coming from quantum gravity corrections to the flat spacetime. Then we are trying to describe a curved spacetime, with constant curvature $\Lambda$, seen by an observer will finite resolution $\delta l$ as a flat spacetime.

Now, assuming the observer is working in a flat spacetime, small movements around him can not be described infinitesimally but are necessarily larger than $\delta l$. This finite displacements are embedded in the original curved spacetime: the "tangent space" to our effective flat spacetime is curved. Thus we get a curved momentum space on a flat spacetime. The curvature of the momentum space is related to a maximal moment norm, which is given by the minimal length scale $\delta l$ on the original manifold. On the other hand, the maximal length scale $l_C$ on the original spacetime defines a minimal moment scale.
Putting the right units where they belong, we start with a curved spacetime with curvature $\Lambda$ seen with a resolution $\delta l$ provided with the usual flat tangent space, and we show it is equivalent to a flat spacetime but provided with a curved moment space with curvature $(\delta l/\hbar)^2$ and a resolution $\delta p\equiv \hbar\sqrt{\Lambda}$.

Considering that $\delta l \ll l_C$, we can work in the limit $\Lambda\arr 0$ , or precisely $z\arr 0$. Then we are the mathematical structure of a smooth curved "tangent space". The coordinates on the flat spacetime need to be reconstructed as the translation generators on the moment space. Now the moment space is curved, so the coordinate becomes non-commutative: 
\be
[X_\mu,X_\nu]=-\f{i \hbar^2}{\kappa^2}J_{\mu\nu}=-i(\delta l)^2J_{\mu\nu},
\ee
where we have defined the quantum deformation parameter $\kappa=\hbar/\delta l$. Mathematically, the moment space is identified as the hyperboloid $SO(4,1)/SO(3,1)$ and the coordinate operators are the DeSitter translation generators.
And we recover the mathematical set-up of deformed special relativity (DSR).

Diagonalizing the coordinate operators, we get a discrete space structure and a continuous time. Another choice would be to consider an AntiDeSitter space, which would lead to the reverse structure of a continuous space and a discrete time. However, one can always question the experimental relevance of such spectra. Indeed, a quantized time would necessary implied phenomenologically a discrete space structure since we traditionally use time-of-flight experiments to measure spacelike distances. Moreover, any clock used to measure time intervals will have discrete ticks and time will always be operationally quantized.

The (squared) length is still expressed as $L^2=X_\mu X^\mu$ and is a Lorentz-invariant: we are truly working in a flat spacetime with a trivial metric. The resulting uncertainty relations read as:
$$
\delta X_\mu\, . \,\delta X_\nu=(\delta l)^2 \la J_{\mu\nu}\ra,
$$
so that $\delta X\sim \delta l$, which is consistent with the starting point that the observer has a finite resolution $\delta l$ on measurements on the space-time coordinates.

It is possible to generalize this construction to an arbitrary initial metric, which gets translated into an arbitrary metric on the moment space, and thus take into account some dynamical effects of quantum gravity \cite{dsreffect}.

Furthermore, one doesn't need to take the limit $\Lambda\arr 0$. Then one needs to take into account the finite resolution of the momentum space which gets translated into a commutation relation:
\be
[p_\mu,p_\nu]=-i(\delta p)^2J_{\mu\nu}.
\ee
This leads to a doubly deformed special relativity. Although it is known that the deformed special relativity algebra is isomorphic to the $\kappa$-deformation of the Poincar\'e group, it is only conjectured that the algebra behind this extension to $\Lambda\ne 0$ is isomorphic to the quantum deformation $\so_q(4,1)$ with parameter $q=e^{-z}$.

\medskip

Now that we have shown how the non-commutative spacetime of deformed special relativity provides a motivated arena to study effective quantum gravity, the issue is to make sense of the DSR physics and check whether we can formulate a consistent kinematics. The main obstacle to a straightforward interpretation of DSR is the "soccer ball" problem: $\kappa$ defines a mass scale which actually becomes a maximal mass when looking at the dispersion relation on the theory. Intuitively, it is the same thing as when we deformed the flat velocity space into the mass-shell hyperboloid in special relativity in order to introduce a maximal speed $c$ \cite{sr}: deforming the moment space into a 4-hyperboloid introduces the maximal rest mass $\kappa$. However the existence of a maximal rest mass is not consistent with the existence of macroscopic system. Indeed in the limit case where $\delta l=l_P$, $\kappa$ is the Planck mass and a soccer ball weights more than $m_P$ for instance.

The solution is to consider that one copy of the algebra describes a "fundamental" system or one-particle system for the observer. To describe a two-particle states, we would need two copies of the DSR algebras. Then the algebra describing the kinematics of the center of mass would be the diagonal part of the direct sum of these two algebras. It turns out that if the two original algebras have $\kappa$ as deformation parameter, then the center of the mass algebra has $2\kappa$ as deformation parameter \cite{dsr}. Thus, let us start at the Planck length with the Planck mass bound: considering systems larger than the Planck length, i.e. many-particle states, will require to renormalize the mass bound $\kappa$ from the Planck mass $m_P$ to the actual maximal mass $M_{max}(L)$ corresponding to the length scale of the system.

Following this logic, \cite{dsr} proposes a law of addition of momenta using a five-component energy-momentum and a generalized relativity principle which states that this new extended 5-momentum is conserved during scattering processes in any reference frames. The deformation parameter $\kappa$ then becomes an intrinsic property of the considered system, in some sense the maximal mass for the system could have. In this effective theory context, the semi-classical limit is then defined as the $\kappa\arr\infty$ limit, when DSR effects become negligible.

\section{Localizing Points in Non-Commutative Spaces}

As we have seen in the previous section, non-commutative geometries are a natural set-up for effective theories of quantum gravity. In fact, already special relativity can be considered as a non-commutative geometry. Indeed the curved space of velocities induces the non-commutativity of the space coordinates if we look at special relativity from a purely 3d space point of view. More precisely, for a given massive system, \cite{sr} derives the following commutation relation:
$$
[x_i,x_j]\,=\,i\,l_c^2\,J_{ij},
$$
where $l_c=\hbar/mc$ is the Compton length associated to the system.

More generally, as argued in \cite{discrete}, requiring Lorentz invariance at the same time as some discrete  spacetime structure leads naturally to geometry (length and time) been defined as (quantum) operators. In such a framework, a necessary step towards talking about a semi-classical limit with a smooth spacetime manifold is to introduce a notion of localized spacetime points: it is necessary to build coherent states of geometry describing a notion of "semi-classical points".

Let us look at the simplest case: we consider 3d space coordinates forming a $\su(2)$ algebra. So we define $x_i=l_P\,J_i$ with $[J_i,J_j]=i\epsilon_{ijk}J_k$. This can actually be seen as 3d Loop Quantum Gravity. The length spectrum is obviously derived:
\be
\hat{l}^2\,=\,l_P^2\,\vec{J}^2
\quad\Rightarrow\quad
l^2\,=\,l_P^2\,j(j+1).
\ee
Then one naturally defines the spread of geometry states as the uncertainty
$(\delta l)^2=|\la\vec{x}^2\ra-\la\vec{x}\ra^2|$. Evaluating it on the $J_z$ eigenvectors $|j m\ra$, we have:
\be
(\delta l)_{|jm\ra}\,=\,l_P\,\sqrt{j(j+1)-m^2}.
\ee
It is easy to show that the minimal spread is obtained for the maximal weight vector $m=j$, for which we have:
$(\delta l)_{min}=l_P\sqrt{j}$. These are the usual $SU(2)$ coherent states, as defined by Peremolov\cite{cohstates}, and provide the notion of semi-classical points with minimal spread.
Let us conclude this example by pointing out that we derive here the uncertainty relation:
\be
\delta l\,=\, \sqrt{ll_P},
\ee
which matches the prediction from distance measurement uncertainty as derived from the holographic principle in three spacetime dimensions (see \cite{discrete} and references therein).

We hope that such a study of semi-classical properties of non-commutative spacetime can be extended to the four spacetime dimension case, in which the holographic principle would require  $(\delta l)^3\,=\, ll_P^2$.

\section{Quantum Geometry from Quantum Information?}

A field which becomes more and more relevant to the study of quantum gravity is quantum information. It has been often proposed that the quantum geometry be reconstructed in terms of flow of information at the fundamental level, especially since the holographic principle has been put forward as a fundamental principle for quantum gravity.
Let us start by a simple remark about the precision of measurements. If one would like increase the precision of the measurement of a property of a system, one would need to use a larger and more massive measurement device in order to achieve the macroscopic amplifications of the smaller fluctuations. More example, the precision bound on a length measurement goes as $\delta l\sim\hbar/cM$. But changing the measurement device to a larger one will modify the geometry of the surrounding and perturb the measured system. Of course, one could argue that we could place the measurement device further in order to lessen the perturbation of the geometry that it causes. Nevertheless, ultimately, the information it self which we extract from the measured system should carry some energy and some mass which perturb itself the geometry and the system. Thus, they should exist a fundamental uncertainty principle between the information extracted on a system (or the geometry of a space-time region) and the reliability (or truth) of this information (or equivalently the perturbation of the system or geometry).
The semi-classical regime could then be defined as minimizing such an uncertainty relation. 

A quantum information concept of particularly interesting for quantum geometry is the notion of quantum reference frames. Indeed, these are quantum systems who state spaces approximates a classical symmetry group. When the symmetry group is the 3d rotation group or the Lorentz group, then we have a quantum system which can be used to localize a direction. Imagining a system composed of many such quantum reference frames would make a reasonable quantum geometry background. More precisely, the main difficulty in diffeomorphism invariant theories such as general relativity is to construct relevant physical observables. The practical solution is to use relational observables, built as physical correlations between two systems $A$ and $B$: we are dealing with the probability amplitude of $A$ being in some state when $B$ is in such or such a state. Choosing $B$ to be given by one of these quantum reference frame would allow us to "localize" the system $A$, with respect to the reference system $B$. Imagining a quantum geometry region made of many quantum reference frames would allow us to have a notion of "where", which is necessary to talk about a semi-classical geometry.

Finally, in a background independent quantum gravity theory, there is no natural notion of distance. Considering a state of quantum geometry of, we should first be able to define a subsystem of this geometry, which would correspond to the notion of a region of space(time). Then in a diffeomorphism invariant context, the only notion of distance in a purely geometrical theory which can be natural is given by the simple criteria: if two systems are close, then they are strongly correlated;  the furthest they are, the weaker the correlations are. More precisely, we could define a distance in terms of correlation and entanglement. I propose to talk about a semi-classical geometry state when this notion of distance is well-behaved and continuous.

To conclude this section, we see that there is a lot of room for an interesting interplay between quantum information
and research in quantum gravity \cite{ohlqg}. All the previous remarks can be applied to loop quantum gravity, which is the main candidate to a theory realizing the quantization of general relativity.

\section{Semi-Classical (Loop) Quantum Gravity}

Loop quantum gravity\cite{lqg1,lqg2,lqg3} is a canonical quantization of general relativity. It describes the quantum states of space geometry and their evolution in time. These are the spin networks, graph decorated with $SU(2)$ spin representations on their edges (a half-integer number $j\in\N$) and with $SU(2)$ invariant tensors at their vertices (Clebsh-Gordan coefficients). The main result of this quantization is the derivation of a discrete spectrum for the area and volume 
operators. More precisely, a spin network can be understood as representing a discrete manifold. Vertices represent chunks of 3d volume. The graph links are dual to the surfaces boundary of these pieces of volume. An elementary surface intersecting only one link of the spin network labeled by a spin $j$ has an area $j\,l_P^2$. Actually, there is an ambiguity in the area spectrum, which could also be $\sqrt{j(j+1)}$ or $j+1/2$, but these details do not matter in the regime of large spins $j$. A generic surface is decomposed in elementary patches, each of them intersecting the spin network only once. The total area of the surface is the sum of the area of all its elementary patches. More precisely, noting $\Gamma$ the spin network and ${\cal S}$ the surface, the area is:
\be
{\cal A}({\cal S})\,\equiv\,l_P^2\sum_{M\in{\cal S}\cap\Gamma}a(M),
\ee
where $a(M)$ is the elementary area, $j$ or $\sqrt{j(j+1)}$, associated to the intersected spin network edge.

In this context, I could talk about the construction of quantum reference frames as discussed in \cite{ohlqg}. But I prefer to discuss the renormalisation of geometrical quantities. Let us consider the area of a surface. The actual measure area will depend on the resolution of the observer doing the measurement. Just like for fractals, the  measured macroscopic area will depend on the way the surface is folded in space. The surface is made of elementary patches, each of them is attached a $SU(2)$ spin representation. The way the surface is folded is described through a tensor between these $SU(2)$ representations. If the surface is closed, it should be an invariant tensor -an intertwiner. This tensor or intertwiner defines the way the surface will get coarse-grained. And ultimately, one could coarse-grain it to a single elementary patch, which represents the surface macroscopically. One can prove that the most probable macroscopic area goes as the squareroot of the microscopic area when we have no knowledge of the way the surface is folded:
$$
\textrm{trivial geometry state}
\Rightarrow
{\cal A}_{macro}\sim l_P\sqrt{{\cal A}_{micro}}.
$$
Knowledge will influence the coarse-graining flow of the surface. Now one should understand that that knowledge of the tensor or intertwiner between the surface elementary patches is actually given by the spin network, which defines the way the surface if embedded in the surrounding volume.

Reversing this logic, describing the renormalization flow of the area of surfaces can be considered as providing the whole quantum state of geometry. Already at a naive level, one can see that the sign of the curvature is related to the increasing or decreasing of the flow. It should nevertheless be investigated in details whether one could actually reconstruct the underlying spin network state or not. 

To sum up, we can trade the knowledge of the state of quantum geometry for the information about the coarse-graining flow of areas of surfaces in loop quantum gravity. This actually makes sense in such a diffeomorphism invariant theory: as one needs to talk about an abstract surface since we can not localize it, we can at least talk about the renormalisation flow of its geometrical properties. We can push this further and directly define a quantum geometry state as the renormalisation flow of the areas in loop quantum gravity. More precisely, as the probability amplitude distribution of the tensors or intertwiners describing the way the surface is folded and thus the way it gets coarse-grained. From this point of view, each quantum geometry state contains the information which defines its continuum limit or semi-classical limit.


\medskip

To conclude, the notion of semi-classicality is an important issue in quantum gravity. In a background independent theory, either one could introduce effective description of quantum gravity depending explicitly on the observer and taking place in a (non-commutative) flat spacetime, or one should introduce new ways to define the semi-classical regime in a background independent fashion. I proposed to address the latter alternative using tools of quantum information and developing the mathematical relation between the quantum state of geometry and the corresponding renormalisation/coarse-graining of geometrical objects.

\section*{Acknowledgments}

I am grateful to Laurent Freidel, Florian Girelli and Daniele Oriti for plenty of discussions on quantum gravity and deformed special relativity.

\bibliography{apssamp}

\end{document}